\documentclass[aps,prd,twocolumn,groupedaddress,showpacs]{revtex4}

\usepackage{graphicx}
\usepackage{graphicx}
\usepackage{amssymb}
\usepackage{amsmath}
\usepackage{color}
\newcommand{\be}{\begin{equation}}
\newcommand{\ee}{\end{equation}}
 
 \definecolor{BrickRed}{cmyk}{0,0.89,0.94,0.28}
\definecolor{MidnightBlue}{cmyk}{0.98,0.13,0,0.43}
\definecolor{DarkGreen}{rgb}{0,0.7,0.1}

\begin{document}

\title{Beyond-PFA Casimir Force between two  spheres at finite temperature.}


\author{Giuseppe Bimonte}

\affiliation{ Dipartimento di Fisica E. Pancini, Universit\`{a} di
Napoli Federico II, Complesso Universitario
di Monte S. Angelo,  Via Cintia, I-80126 Napoli, Italy}
\affiliation{INFN Sezione di Napoli, I-80126 Napoli, Italy}

\email{giuseppe.bimonte@na.infn.it}

\begin{abstract}

A recent experiment  [J.L. Garrett et al., Phys. Rev. Lett {\bf 120}, 040401 (2018)] measured for the first time the gradient of the Casimir force  between two gold spheres at room temperature. The theoretical analysis of the data was carried out  using the standard Proximity Force Approximation (PFA).  A fit  of the data, using a parametrization of the force valid for the sphere-plate geometry, was used by the authors to place  a bound on deviations from PFA.  Motivated by this work,  we   compute the Casimir force between two gold spheres at finite temperature.  The semi-analytic formula for the Casimir force that we construct  is valid for all separations, and  can be easily used to interpret future experiments in both the sphere-plate and sphere-sphere configurations. We describe the correct parametrization of the corrections to PFA for two spheres that should be used in data analysis.

\end{abstract}


\maketitle

\section{Introduction}
\label{sec:intro}

The Casimir force \cite{Casimir48} is the tiny long-range force between (neutral) macroscopic polarizable bodies that originates from the modification of the spectrum of quantum and thermal vacuum fluctuations of the electromagnetic  (em) field,  caused by the presence of the bodies. This phenomenon represents one of the rare manifestations of the quantum properties of the em field at the macroscopic scale. For  reviews   see Refs.\cite{book1,parse,book2,woods,mehran}.

A characteristic feature of the Casimir force is its {\it non-additivity}, reflecting the   many-body character of fluctuation forces. This property  enormously complicates the  computation the Casimir force in  non-planar geometries. We recall that in his original paper Casimir worked out the force in the highly idealized geometry of two perfectly conducting plane-parallel surfaces at zero temperature.   Planar systems were the exclusive object of consideration also in the famous paper by Lifshitz \cite{lifs},   who derived a general formula for the Casimir interaction between two plane-parallel slabs, taking into full account  realistic material properties of the plates, i.e. their frequency-dependent dielectric permittivity and finite temperature.

Unfortunately, the theoretically simple planar geometry studied by Casimir and Lifshitz has been rarely used in experiments \cite{sparnaay, gianni}, because of severe difficulties connected with controlling the parallelism of two macroscopic surfaces separated by a submicron gap.   To avoid these problems, the vast majority of Casimir experiments adopt the sphere-plate geometry (see for example  \cite{lamor1,umar,semic,decca6,liq,ito,RSDPalasantzas2010,lamorth,chang2,bani2,deccamag} and Refs. therein), which is obviously immune from parallelism issues.  

Very recently, a new experiment   \cite{garrett}   has measured for the first time the (gradient of the) Casimir force   between two gold-coated spheres.  Following the practice of previous experiments, also in this new experiment the Casimir force has been computed using  the simple Proximity Force Approximation (PFA) \cite{Derjaguin}, which expresses the Casimir force between two curved surfaces as the average of the plane-parallel force (as given by Lifshitz formula)  over the local separation between the surfaces.  The results of the experiment have been  found to be in good agreement with theoretical predictions based on Lifshitz formula.  By using  the measurements made with nine sphere-sphere and three sphere-plate systems of different radii,  the authors of  \cite{garrett} could also place a bound  on the magnitude of  deviations from PFA,  using the same parametrization of the force that was used in the sphere-plate experiment of the IUPUI group  \cite{ricardo}. 

Motivated by the new experiment, in this paper we compute beyond-PFA curvature corrections to the Casimir force gradient for   two gold spheres at room temperature, and we describe the correct parametrization of the force that should be used in the data analysis to measure corrections to PFA in this geometry.  We remind the reader that the computation of the Casimir force for non-planar  geometries has   been an untractable problem until recently.  Only in the early 2000's an exact scattering formula for the Casimir interaction between  dielectric objects of any shape, generalizing   early results of Balian and Duplantier \cite{Balian}  and Langbein \cite{langbein}, has been worked out  \cite{sca1,sca2, kenneth}. At finite temperature $T$, the scattering formula has the form of a sum over so-called Matsubara (imaginary) frequencies $\xi_n= 2 \pi n k_B T/\hbar, \;n=0,1,\dots$ (with $k_B$ Boltzmann constant, and $\hbar$ Planck constant) of functional determinants involving the multipole expansions of the  T-operators of the two bodies. Unfortunately, the scattering formula converges very slowly in the characteristic regime of Casimir experiments, in which the (minimum) separation $a$ between the  two surfaces  is very small compared to their characteristic radius of curvature $R$.  In order to obtain a precise estimate of the Casimir force, as it is needed  for a proper interpretation of current precision experiments,  it is necessary to push the computation  to very high multipole orders, which represents a very challenging task even for present day computers.  In \cite{antoine1} simulations of the sphere-plate problem for Drude conductors were done up to  multipole order $l_{\rm max}=24$, which were later \cite{antoine2}  pushed to $l_{\rm max}=45$, allowing to calculate the force for $a/R \ge 0.1$. Very recently \cite{gert}, a  large-scale numerical simulation of the sphere-plate system going up to mutipole order $l_{\rm max}=2 \times 10^4$   reached for the first time the experimentally important region $a/R \sim 10^{-3}$.

Large numerical simulations of the scattering formula like that of \cite{gert} require sophisticated algorithms for the computation of determinants of hierarchical matrices, that non every researcher may master or be willing to spend time on. This led us to investigate if the new  analytical tools that have been developed recently in the Casimir field,  could be exploited to construct easy-to-use formulae for the Casimir force having the high degree of precision demanded by current experiments.
In Ref. \cite{bimonteprecise}   a formula with these features was constructed for the sphere-plate system. 

The  approach followed in \cite{bimonteprecise} can be described as follows.  As we said earlier, the scattering formula has the form of a sum over  Matsubara modes $\xi_n$. The first term of this series, corresponding to $n=0$, represents a classical contribution to the Casimir interaction, which becomes dominant in the high temperature limit $a/\lambda_T \gg 1$, where $\lambda_T=\hbar c/2 \pi k_B T$ ($\lambda_T=1.2 \;\mu$m for room temperature)  is the thermal length. In Ref. \cite{bimonteex1},  it was shown that this classical term can be evaluated {\it exactly}  in the geometry of two metallic spheres of any radii, including the sphere-plate case as a special limit.  Of course, knowledge of the $n=0$ term, is not sufficient to compute the full Casimir energy. Unfortunately, the   $n>0$ terms of the scattering formula cannot be  computed exactly, but in \cite{bimonteprecise}  it was shown that they can be computed very precisely using an  asymptotic small-distance formula,  which includes corrections to  PFA,    based on a recently proposed Derivative Expansion (DE) of the Casimir interaction \cite{fosco1,bimonte1,fosco2,bimonte2,fosco3}. 
 In  \cite{bimonteprecise} it was proved that the semi-analytic approximate formula  for the sphere-plate Casimir force, resulting from the combination of  the exact $n=0$ term with the approximate expression of the $n>0$ terms,  is indeed  extremely precise  {\it for all separations}. The   formula derived in \cite{bimonteprecise} is in excellent agreement with the results of the large numerical simulation of \cite{gert}.  In this paper, we extend the construction  of \cite{bimonteprecise} to the sphere-sphere system.

The paper is organized as follows: in Sec. II we review the PFA for two spheres,  the scattering formula
and display the exact  solution   for the classical Casimir energy of two Drude spheres  discovered in \cite{bimonteex1}. In Sec. III we compute the contribution of the positive  Matsubara modes using the DE.   In Sec. IV we display our complete formula for the Casimir energy of two  spheres at finite temperature, and use it to compute deviations from PFA. In Sec. IV we also review the recent two-sphere experiment \cite{garrett} and describe the correct  parametrization of corrections to PFA that should be used  in the data analysis of  experiments with two spheres. In Sec. V we present our conclusions. Finally, in Appendix A we review the DE and in Appendix B we use the DE to compute the leading curvature correction to the force gradient between two spheres.

\section{Casimir interaction of two spheres}

We consider a system composed by two spheres of respective radii $R_1$ and $R_2$ placed in a vacuum and separated by a gap of width $a$ (see Fig. 1). 

As it was explained in the introduction,  until recently there were no tools to exactly compute the Casimir force in non-planar geometries, and so one had to resort to the old-fashioned PFA.  In the case of two spheres
\cite{parse} the PFA formula for the Casimir force is:
\be
F^{(\rm PFA)}(a,R_1,R_2)= 2 \pi {\tilde R} \;{\cal F}^{(\rm pp)}(a)\;,\label{PFA1}
\ee     
where ${\tilde R}$ is the {\it effective} radius of the two spheres
\be
{\tilde R}=\frac{R_1 R_2}{R_1+R_2}\;,
\ee
and  ${\cal F}^{(\rm pp)}(a)$ is the unit-area Casimir free energy for two plane-parallel slabs respectively made of the same materials as the sphere and plate, whose expression was derived long ago by Lifshitz \cite{lifs}:
\begin{eqnarray}
&&
{\cal F}^{(\rm pp)}(a,T)=\frac{k_BT}{2\pi} \sum_{n \ge 0}\;\!\!'
 \int_{0}^{\infty}
k_{\bot}dk_{\bot} \nonumber
\\
&&\times
\sum\limits_{\alpha={\rm
TE,TM}}\ln\left[1-
{r_{\alpha}^2({\rm i}\,\xi_n,k_{\bot})}e^{-2aq_n}\right],\label{PBeq3}
\end{eqnarray}
where the prime in the sum over $n$ indicates that the $n=0$ term is taken with a weight 1/2,  $T$ is the temperature of the plates, $k_{\bot}$ is the in-plane momentum, $r_{\alpha}({\rm i}\,\xi_n,k_{\bot})$ denotes the Fresnel reflection coefficient for polarization $\alpha={\rm TE, TM}$ of a thick slab, evaluated for the imaginary frequency $\omega={\rm i}\, \xi_n$ and 
$q_n=\sqrt{\xi_n^2/c^2+k_{\bot}^2}$. The PFA for the force gradient $F' \equiv \partial F/\partial a$  (here and in what follows a  prime shall denote a derivative with respect to the separation) easily follows from Eq. (\ref{PFA1}):
\be
{F'}^{(\rm PFA)}(a,R_1,R_2)=- 2 \pi {\tilde R}\; {F}^{(\rm pp)}(a)\;,\label{PFA2}
\ee
where ${F}^{(\rm pp)}(a)=-\partial  {\cal F}^{(\rm pp)}/\partial a$ is the unit-area Casimir force for two parallel slabs. The  PFA  force and force-gradient  for a  sphere of radius $R$ opposite a plane, is recovered from Eqs. (\ref{PFA1}) and (\ref{PFA2}), respectively, by taking the radius of one of the two spheres to infinity, i.e. substituting  ${\tilde R}$ by $R$. It is important to remark that within the PFA, both $F$ and $F'$ depend on the radii of the two spheres only via the effective radius ${\tilde R}$.
 
For a proper interpretation of current precision Casimir experiments it has become important to estimate curvature corrections beyond PFA. This has been impossible until recently, when 
an exact scattering formula providing the Casimir energy  of two compact dielectric bodies has been worked out \cite{sca1,sca2,kenneth}.  The general structure of the scattering formula is:
\be
{\cal F}=k_B T \sum_{n \ge 0}\;\!\!' \;{\rm Tr} \ln[1-\hat{M}(\rm {i} \xi_n)]\;,\label{freeen}
\ee  
where the prime sign in the sum indicates again that the $n=0$ term is taken with weight 1/2.  The trace ${\rm Tr}$ in this equation is  over both spherical multipoles indices $(l,m)$ and   polarization indices $\alpha={\rm TE,TM}$:
\be
{\rm Tr}= \sum_{m=-\infty}^{\infty} \sum_{l=  |m|}^{\infty} {\rm tr}\;,
\ee  
where ${\rm tr}$ denotes the trace over $\alpha$. The matrix elements  $M_{lm\alpha,m'l'\alpha'}$ of ${\hat M}$ shall not be reported here for brevity. Their explicit expressions can be found for example in Refs.\cite{rahi,teo2}. We just recall that the matrix $M_{lm\alpha,m'l'\alpha'}$ involves a product of the T-matrices for the two bodies (i.e. the Mie scattering coefficients in the case of  two spheres), both evaluated for the imaginary Matsubara frequencies ${\rm i} \xi_n$,  intertwined  with  translation matrices that serve to convert the mutipole basis relative to either body into the multipole basis relative to the other body (see Refs.\cite{rahi,teo2} for details). The expressions for the Casimir force $F=-  {\cal F}'$ and its gradient $F'$  are obtained by taking  derivatives of Eq. (\ref{freeen}) with respect to the separation $a$. Using the scattering formula  it has been possible to   prove eventually that the PFA formula is indeed asymptotically exact for small separation in the sphere-plate and cylinder-plate geometries  \cite{bordag}. 

Having at our disposal the exact representation  Eq. (\ref{freeen}), it is natural to ask whether it can be used efficiently to accurately compute the Casimir force in concrete experimental situations. Unfortunately, this is not easy at all. Consider as an example the geometry of a sphere of radius $R$ at a minimum distance $a$ from a plate. The problem is that to obtain a precise estimate of the Casimir force for experimentally relevant sphere-plate separations (typical aspect ratios $a/R \sim 10^{-3}$) it is necessary to include a huge number of multipoles in the computation. Previous
works  \cite{sca2,antoine1,antoine2,bimonteprecise,gert} found that the multipole order $l_{\rm max}$ for which convergence  is achieved  scales as $l_{\rm max} \sim R/a$. To date, the largest numerical simulation  of the sphere-plate scattering formula reached up to $l_{\rm max}=2 \times 10^4$   \cite{gert},  which allowed the authors of \cite{gert} to probe the Casimir force in  the experimentally relevant region $a/R \sim   10^{-3}$. Managing such a large number of multipoles on a computer is not easy at all, and sophisticated algorithms are needed to handle the problem. 

\begin{figure}
\includegraphics[width=.9\columnwidth]{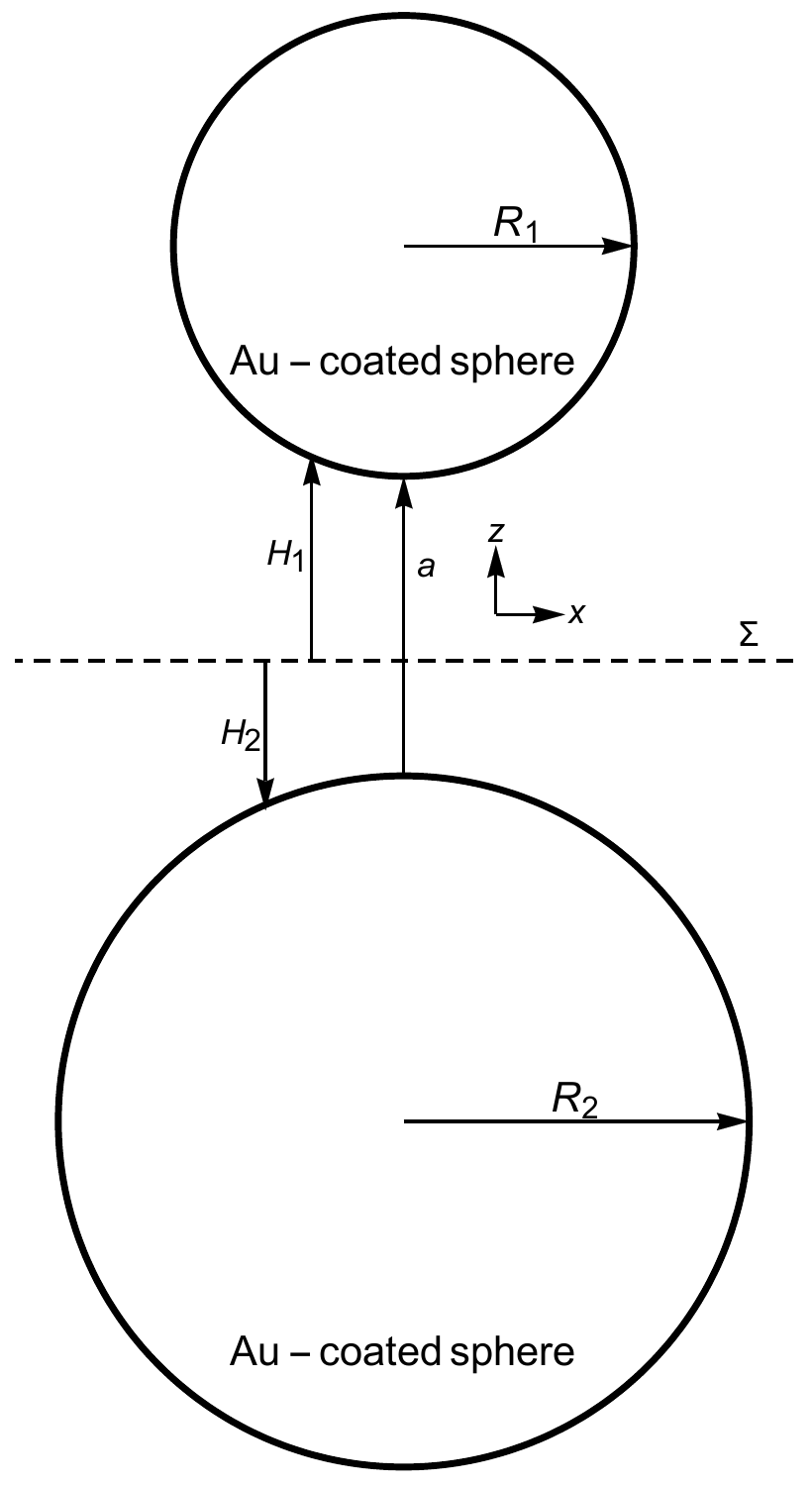}
\caption{\label{setup}  The sphere-sphere Casimir setup. The sphere-sphere geometry is characterized by the {\it effective} radius ${\tilde R}=R_1 R_2/(R_1+R_2)$ and the dimensionless parameter $u={\tilde R}^2/R_1 R_2$}
\end{figure}

At this point we turn to our objective of deriving a simple and very accurate formula for the Casimir force between two spheres. To do this, we go back to the general scattering formula Eq. (\ref{freeen}). As we see, it has the form of a sum of terms ${\cal F}=\sum'_{n \ge 0}{\cal F}_{n}$  over the Matsubara frequencies $\xi_n,\;n=0,1,\dots$. It is convenient for our purposes to separate the first term ${\cal F}_{n=0}$ of the series from the the remaining terms with $n>0$. We accordingly decompose Eq. (\ref{freeen}) as:
\be
{\cal F}={\cal F}_{n=0}+{\cal F}_{n>0}\;,
\ee
where we set ${\cal F}_{n>0}=\sum_{n > 0}{\cal F}_{n}$.
Consider first ${\cal F}_{n=0}$. This term represents a {\it classical} contribution to the Casimir energy, which provides the dominant contribution to the full Casimir energy ${\cal F}$ of the system in the limit of large separations $a \gg \lambda_T$.  

In Ref. \cite{bimonteex1}, it was shown that this classical term can be evaluated {\it exactly}  in  the following two cases. The first one is that  of a scalar field obeying Dirichlet (D) boundary conditions (bc) on the surfaces of two spheres of arbitrary radii, including the sphere-plate geometry as a special case. The second case is that of a scalar field obeying so-called Drude bc on the surfaces of a sphere opposite a plate \footnote{The sphere-plate solution with Drude bc presented in \cite{bimonteex1} does not extend to the sphere-sphere geometry.}. The latter bc is identical to D bc, apart from the  fact that in the Drude case   charge monopoles (corresponding to index $l=0$) are excluded from  the scattering formula. Both sets of bc can be used to describe ohmic conductors, depending on the electric configuration of the system. Drude bc describe isolated conductors, whose total charge is fixed, while D bc  describe conductors whose voltages are fixed. The latter is the experimentally important situation,  since in all Casimir experiments (including the experiment \cite{garrett}) one plate is  grounded, while the other is connected with a voltage generator  which serves to apply a bias potential, to compensate for unavoidable potential differences between the plates, resulting from differences in the respective work functions. The classical sphere-sphere Casimir energies  implied by D and Drude bc  are undistinuishable for separations much smaller than the spheres radii, while  the two models lead to distinct asymptotic behaviors in the limit of large separations $a \gg (R_1,R_2)$, since   ${\cal F}^{\rm (D)}_{\rm n=0} \sim -k_B T R_1 R_2/a^2$, while   ${\cal F}_{\rm n=0}^{(\rm Dr)} \sim -k_B T R_1^3 R_2^3/a^6$. 

For two  metallic spheres obeying D bc the exact solution worked out in \cite{bimonteex1} is:  
\be
{\cal F}_{n=0}^{(\rm ex)}=\frac{k_B T}{2} \sum_{l=0}^{\infty} (2 l +1) \ln[1-Z^{2 l+1}] \; ,\label{enerDr}
\ee
where $Z$ is:
\be
Z=[1+x+x^2 u/2+\sqrt{(x+x^2 u/2)\,(2+x+x^2 u/2)}]^{-1}\;.\label{Zdef}
\ee
In the above Equation, $x=a/\tilde{R}$,  and $u=\tilde{R}^2/R_1 R_2$. The parameter $u$ depends only on the ratio between $R_1$ and $R_2$ and  takes values in the interval $[0,1/4]$,  the upper bound $u=1/4$ corresponding to two equal spheres $R_1=R_2$, while the lower bound $u=0$ is approached as either of the two radii becomes infinite. The  special case of a sphere of radius $R$ opposite a plate is thus recovered by taking $u=0$ into Eq. (\ref{Zdef}), and setting $\tilde{R}=R$.  The  range of the variable $Z$ in Eq. (\ref{Zdef}) is the interval $[0,1]$,  the upper (lower) bound $Z=1$ ($Z=0$) corresponding to the limit of vanishing (infinite) separation $x\rightarrow 0$ ($x\rightarrow \infty$). 
The corresponding expression for the Casimir force $F_{n=0}^{(\rm ex)}= - {{\cal F}'}_{n=0}^{(\rm ex)}$ and its derivative ${F'}_{n=0}^{(\rm ex)}$ are easily obtained by deriving Eq. (\ref{enerDr}) with respect to $a$.  As we see from the expression of $Z$ in Eq. (\ref{Zdef}),  the exact classical energy ${\cal F}_{n=0}^{(\rm ex)}$ depends not only on the effective radius ${\tilde R}$, but also on the ratio among the radii via the variable $u$. This feature marks an important difference with respect to the PFA formula.

In \cite{bimonteex1} the small distance expansion of  ${\cal F}_{n=0}^{(\rm ex)}$ for the sphere-plate system was  worked out, by setting $Z=\exp(-\mu)$ and then taking the small-$\mu$ asymptotic expansion of the series on the r.h.s. of Eq. (\ref{enerDr}). Using the formulae of Ref. \cite{bimonteex1}, it is easy to verify that for small separations  the sphere-sphere  force gradient has the expansion:
\be
 {F'}_{n=0}^{(\rm ex)}=k_B T\frac{ \zeta(3) {\tilde R}}{4\, a^3} \left(1+\frac{1}{12 \zeta(3)} \frac{a}{\tilde R} +o(a/{\tilde R}) \right)\;,\label{DEnzero}
\ee
where   $\zeta(x)$ is Riemann zeta function.
The leading term coincides with the PFA Eq. (\ref{PFA2}), since from Lifshitz formula we find  ${F}^{(\rm pp)}_{n=0}=-k_B T \zeta(3)/(8 \pi a^3)$ for two Drude-metal plates, while the next term provides the leading correction to PFA. Interestingly, like the PFA, also the latter correction is independent of the parameter $u$. The correction to PFA in Eq. (\ref{DEnzero}) is consistent with the DE (see Eq. (\ref{DEformula})).

At this point we need consider the contribution ${\cal F}_{n>0}$ of the positive Matsubara modes $n>0$ to the free energy. Unfortunately, differently from the classical term ${\cal F}_{n=0}$, the quantity ${\cal F}_{n>0}$ cannot be computed exactly. Of course ${\cal F}_{n>0}$ be computed numerically,  using the scattering formula truncated to a finite multipole order $l_{\rm max}$. As we explained earlier, such a computation is however very challenging, because the multipole order $l_{\rm max}$ that is necessary
is very large for  experimentally relevant values of the radii and separation. Below we obtain a very precise and simple analytical formula for ${\cal F}_{n>0}$,  by using the so-called Derivative Expansion  
 \cite{fosco1,bimonte1,fosco2,bimonte2}.

\section{Derivative Expansion of ${\cal F}_{n>0}$}\

The DE \cite{fosco1,bimonte1,fosco2,bimonte2,fosco3} is an analytical technique to compute curvature corrections to proximity forces beween two surfaces of small slope. For the benefit of the reader, we provide in Appendix A a short review to the DE and a guide to the relevant References.

In this Section we use the DE to estimate the contribution ${\cal F}_{n>0}$ of the non-zero Matsubara modes to the Casimir energy. The DE is particularly well suited to this task, as we now explain. By it very nature, the DE is expected to be very precise in situations in which the slope of the surfaces  is small within the interaction region. A little reasoning shows that this condition is met in the problem at hand for all separations $a$ between the spheres, just provided (as it always is the case in current experiments) that the radii   of the spheres are both large compared to the thermal length $\lambda_T$ ($\lambda_T=1.2\;\mu$m at room temperature).   

It is a well known fact \cite{parse,book2} that the Casimir interaction between two surfaces, with a characteristic radius of curvature $R$,  is localized inside a disk of radius $\rho \sim \sqrt{a R}$ around the point of either surface  which is closest to the other. 
Thus, one is led to expect  that the DE is applicable in general only  for  separations $a$ such that $\rho/R=\sqrt{a/R} \ll1$. This condition is usually well satisfied in Casimir experiments, for which typically ${a/R} < 0.01$. A  closer inspection reveals however that the DE of ${\cal F}_{n>0}$   is in fact valid for {\it all} separations, provided that $R_1$ and $R_2$ are both larger than $\lambda_T$. The point to consider is that {\it positive} Matsubara modes of (imaginary) frequency $\xi_n$ can only propagate across a distance
of order $\ell_n=c/\xi_n =\lambda_T/n \le \lambda_T$.  Because of this constraint, the true size of the interaction region is actually not  $\sqrt{a R}$, but instead $\rho=\min (\sqrt{a R},\sqrt{\lambda_T R})$. This implies that  the DE for ${\cal F}_{n>0}$ is actually valid for separations such that $\rho/R=\min (\sqrt{a/R},\sqrt{\lambda_T/R}) \ll 1$. The latter condition is clearly satisfied for all separations, provided that $R_1$ and $R_2$ are both much larger than $\lambda_T$.  

We have thus established that the DE is a valid method for ${\cal F}_{n>0}$. In Appendix B it is shown that the DE leads to a simple general formula Eq. (\ref{DEformula}) for the force gradient between two spheres. The  formula for the DE expansion of  ${\cal F}_{n>0}$ can obtained by making into  Eq. (\ref{DEformula}) the appropriate substitutions:     
\be
{ F}'_{n>0}= -2 \pi {\tilde R} {F}^{(\rm pp)}_{n>0}(a) \left[1- \left(\tilde{\theta}(a)+u \,\kappa(a) \right) \frac{a}{\tilde R} \right]\;, \label{DEformula1}
\ee
where the coefficients $\tilde \theta(a)$ and $\kappa(a)$ are (see Eqs. (\ref{thetacoe}) and (\ref{kappacoe}))
\begin{eqnarray}
{\tilde \theta}&=&  \frac{{{\cal F}}^{(\rm pp)}_{n>0}(a) -2 \alpha_{n>0}(a)}{a {F}^{(\rm pp)}_{n>0}(a)}\;,\label{thetacoe1}\\
\kappa(a) &=&  1-2 \frac{{ {\cal F}}^{(\rm pp)}_{n>0}(a)}{ a { F}^{(\rm pp)}_{n>0}(a)} \;.\label{kappacoe1}
\end{eqnarray}
In the above Equations, $ {\cal F}^{(\rm pp)}_{n>0}$ denotes the contribution of the $n>0$ modes to Lifshitz formula Eq. (\ref{PBeq3}), and ${F}^{(\rm pp)}_{n>0}=-\partial {\cal F}^{(\rm pp)}_{n>0}/\partial a$ is the corresponding (unit area) force.  

An important ingredient of Eqs. (\ref{DEformula1}-\ref{kappacoe1}) is the coefficient $\alpha_{n>0}(a)$ which enters into the expression of $\tilde{\theta}$ in Eq. (\ref{thetacoe1}). As it is explained in Appendix A, this coefficient can be extracted from the Green function  
${\tilde G}^{(2)}(k;a)$ of the second-order perturbative expansion of ${\cal F}_{n>0}$ for a small amplitude deformation of one of the plates around the plane-parallel  geometry. The computation of this coefficient for gold plates at room temperature  can be carried out following the procedure described in \cite{bimonte2}, and we address the interested reader to that Reference for details. The coefficients  $\tilde \theta$ and $\kappa$ for perfect conductors (PC) in the limit of zero temperature are both independent of the separation.  Their values can be determined using the formulae listed in \cite{bimonte1}:  
\be
\tilde \theta^{(\rm PC)}_{T=0}=\frac{20}{3 \pi^2}-\frac{1}{9}=0.564\;,\;\;\;\;\;\;\kappa^{(\rm PC)}_{T=0}=\frac{1}{3}\;.
\ee  For gold surfaces at room temperature, both coefficients depend on the separation (as well as on the temperature and on the material lengths characterizing the optical properties of gold, in particular the plasma length). In Table \ref{tab.1} we list the values of $\tilde \theta(a)$ and $\kappa(a)$ for gold at room temperature, that we computed using tabulated \cite{palik} optical data \footnote{The weighted Kramers-Kronig dispersion relations \cite{bimonteKK} was used to compute precisely $\epsilon({\rm i} \xi_n)$ starting from the real-frequency optical data given by Palik. }, for several values of the separation in the range from 100 nm to 2 micron.  
Using these values of $\tilde \theta(a)$ and $\kappa(a)$ together with Eq. (\ref{DEformula}), it is easily possible to compute  ${ F}'_{n>0}$ for any combinations of sphere radii. 

In view of later applications, it is important to note that while in the Proximity Approximation  ${ F}'_{n>0}$ depends only on the effective radius ${\tilde R}$ (see Eq. (\ref{PFA2})),  the  more accurate expression of ${ F}'_{n>0}$ in Eq. (\ref{DEformula1}), which includes curvature corrections to PFA, depends  also on the parameter $u$, i.e. on the ratio of the radii of the two spheres.
\begin{widetext}
\label{tab.1}
\begin{center}
\begin{table*}
\begin{tabular}{ccccccccccccc} \hline
$a (\mu m)$\;\; &0.10& 0.15 & 0.2  \;\;& 0.25\;\;\;& 0.3\;\; &0.35 \;\;& 0.4 \;\; & 0.45 & 0.5\;\;& 0.55 & 0.6\;\; & 0.65  \\ \hline \hline
${\tilde \theta}$\;\;& 0.456  &0.4715 & 0.470\;\; &0.463 & 0.454 \;\;&0.4445 & 0.435 \;&0.425 & 0.415 &0.4055 & 0.396& 0.387    \\  \hline
${\kappa}$\;\;&0.245 &0.270 &\; 0.289\;\; & 0.305 & 0.319 \;\;&0.331 & 0.342\;\;&0.353  & 0.362\;\;&0.371 & 0.380\;\; &0.389 \;\;    \\ \hline \\
 \hline
$a (\mu m)$\; &0.70&0.75 & 0.8 & 0.85 & 0.9& 0.95&  1 &1.2&1.4 &1.6 & 1.8 & 2  \\ \hline \hline
${\tilde \theta}$ \;\; &0.379 &0.370  &0.362 &0.3545 &0.347&0.3395&0.332 &0.306&0.282&0.261& 0.242 &0.225   \\  \hline
${\kappa}\;\;$&0.397&\;0.405 \;&0.413 &0.421&0.429&0.437& 0.444\;\;&0.474&0.502&0.529&0.554&0.578 \\ \hline  \hline
\end{tabular}
\caption{Values of the coefficients ${\tilde \theta}$ and $\kappa$ for Au at room temperature.}
\end{table*}
\end{center}
\end{widetext}

\section{Beyond-PFA corrections.}

Combining Eq. (\ref{enerDr}) with Eq. (\ref{DEformula1}) we obtain the following formula for the force gradient between two gold spheres:
\be
F'=  {F'}_{n=0}^{(\rm ex)}  -2 \pi {\tilde R} {F}^{(\rm pp)}_{n>0}(a) \left[1- \left(\tilde{\theta}(a)+u\, \kappa(a)\right) \frac{a}{\tilde R} \right]\;,\label{approx}
\ee
which constitutes the main result of the present work. A nice feature of the formula above is that, by construction,  it is {\it exact in both limits} $a/{\tilde R}\rightarrow 0$ and $a/\lambda_T\rightarrow \infty$. This is so because, on one hand, Eq. (\ref{approx}) is exact  for $a/{\tilde R}\rightarrow 0$, since in this limit the DE, which we used to compute the contribution of the positive Matsubara modes,  is asymptotically exact. On the other hand, Eq. (\ref{approx}) is exact  also for $a/\lambda_T \rightarrow \infty$, because for separations larger than the thermal length $a \gg \lambda_T$ the relative contribution of the positive  Matsubara modes vanishes exponentially fast,  and then Eq. (\ref{approx}) reduces to the exact  $n=0$ mode. Comparison with high precision numerical computations of the sphere-plate scattering formula in \cite{bimonteprecise} revealed that Eq. (\ref{approx}) is in fact very accurate also for {\it all intermediate} separations.  For a gold sphere as small as 8 micron, the maximum error made by Eq. (\ref{approx})  was only of 0.1$\%$, and this was for the large aspect ratio $a/R=0.12$. The error  is expected to be far smaller in the conditions of the experimet \cite{garrett}, which used spheres with radii larger than 29.8 micron, and probed distances corresponding to aspect ratios smaller than 0.017.

For later use, it is useful to work out the small distance limit of our formula for the sphere-sphere force gradient. This can be easily done subsituting  $ {F'}_{n=0}^{(\rm ex)} $ on the r.h.s. of Eq. (\ref{approx}) by its small-distance expansion Eq. (\ref{DEnzero}). After simple algebraic manipulations, one finds:
\be
F'=-2 \pi {\tilde R} {F}^{(\rm pp)}(a) \left[1- \left(\hat{\theta}(a)+u\, \hat{\kappa}(a)\right) \frac{a}{\tilde R}+ o(a/{\tilde R}) \right]\;.\label{DE2s}
\ee
 The coefficients $\hat{\theta}(a)$ and $\hat{\kappa}(a)$ are 

\begin{eqnarray}
\hat{\theta}(a)&=&\frac{ {F}^{(\rm pp)}_{n>0}}{{F}^{(\rm pp)}} \; \tilde{\theta}-\frac{{F}^{(\rm pp)}_{n=0}}{{F}^{(\rm pp)}} \,\frac{1}{12\, \zeta(3)}\;,\nonumber \\
\hat{\kappa}(a)&=&\frac{{F}^{(\rm pp)}_{n>0}}{{F}^{(\rm pp)}}\;{\tilde \kappa}\;.\\
\end{eqnarray}

\begin{figure}
\includegraphics[width=.9\columnwidth]{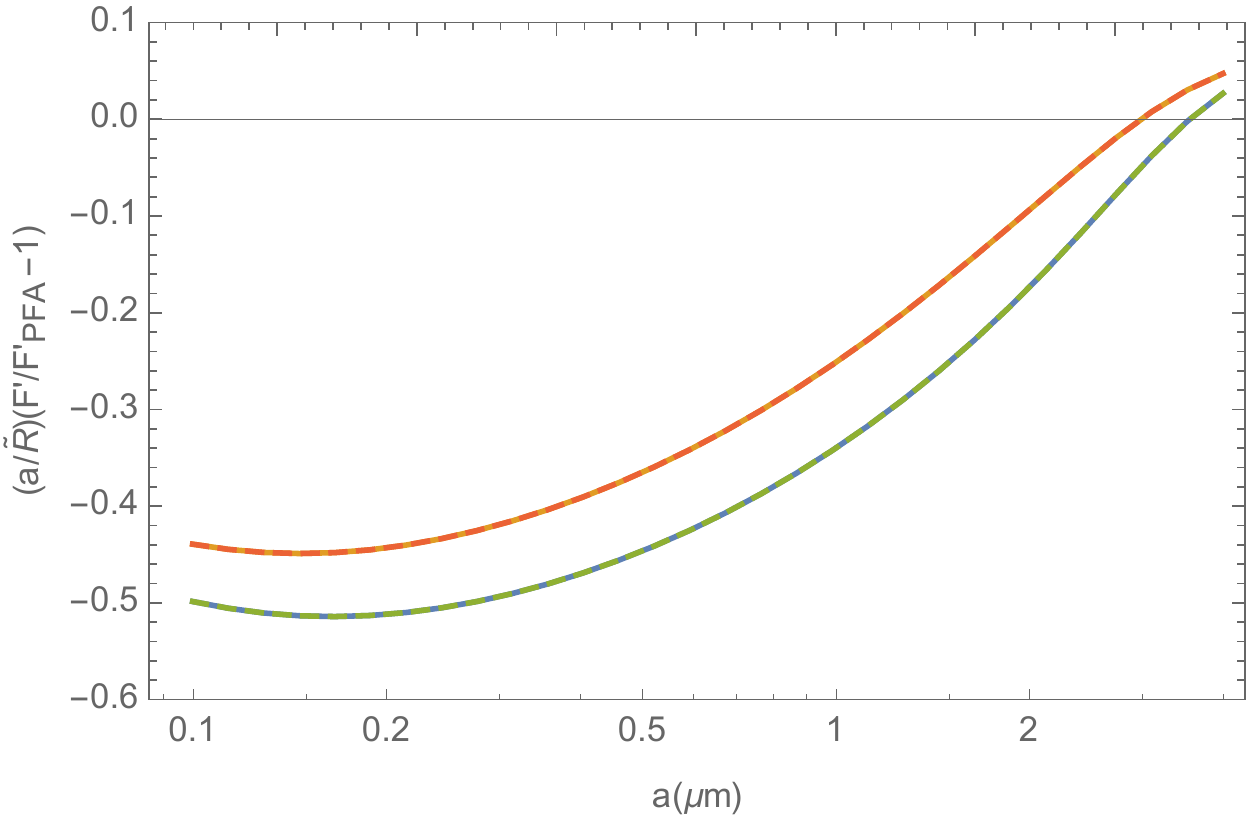}
\caption{\label{deviations}   Beyond-PFA corrections for the gradient of the Casimir force between two gold spheres at room temperature are shown as a function of the separation for two identical spheres of radius $R=30\;\mu$m (lower solid line) and $R=100\;\mu$m (lower dashed line). The upper pair of lines is for a sphere-plate system with sphere radius   $R=30\;\mu$m (upper solid line) and $R=100\;\mu$m (upper dashed line). }
\end{figure}

In Table \ref{tab.2}, we provide the values of the coefficients ${\hat \theta}$ and ${\hat \kappa}$ for gold at room temperature. We note that the coefficient ${\hat \theta}$ introduced here coincides with the opposite of the coefficient ${\hat \theta}_1$ of \cite{bimonte2}.  
\begin{widetext}
\label{tab.2}
\begin{center}
\begin{table*}
\begin{tabular}{ccccccccccccccc} \hline
$a (\mu m)$\;\;&0.05 &\;0.10 & 0.15 & 0.2  \;\;& 0.25\;\;\;& 0.3\;\; &0.35 \;\;& 0.4 & 0.45 & 0.5 & 0.55 & 0.6 \\ \hline \hline
${\hat \theta}$\;\;& 0.378  &\;\;0.439 &\; 0.449\;\; &\;0.443\;\; &\; 0.432 \;\;&0.419 & \;0.405 \;&\;0.392&\;\; 0.378 &\;\;0.365 & \;\;0.352 &\;\;0.340      \\  \hline
${\hat \kappa}$\;\;&0.209 &\;\;0.237 &\; 0.259\;\; & 0.275 &\; 0.288 \;\;&0.298 &\; 0.306\;\;&\;0.313 &\;\;\;0.320 &\;\;0.325 &\;\;0.330 &\;\; 0.334  \\
  \hline  \hline
\end{tabular}
\caption{Values of the coefficients ${\hat \theta}$ and ${\hat \kappa}$ for Au at room temperature.}
\end{table*}
\end{center}
\end{widetext}
In order to show   the deviations of the force gradient  from PFA predicted by our formula, in Fig. \ref{deviations} we  plot the quantity  $({\tilde R}/a)(F'/F'_{\rm PFA}-1)$  for a system of two identical spheres of radius $R_1=R_2=30\;\mu$m (lower solid line) and $R_1=R_2=100\;\mu$m (lower dashed line).
The upper solid and dashed  lines in Fig. \ref{deviations} refer to a sphere-plate system, for  a sfere of radius $R=30\;\mu$m (solid line) and $R=100\;\mu$m (dashed line).  Fig. \ref{deviations}  demonstrates that    deviations from PFA of the force gradient  are practically independent of the effective radius $\tilde R$, for  fixed value of $u$. However deviations from PFA  depend significantly on the ratio among the radii of the spheres via the parameter $u$.  
\begin{figure}
\includegraphics[width=.9\columnwidth]{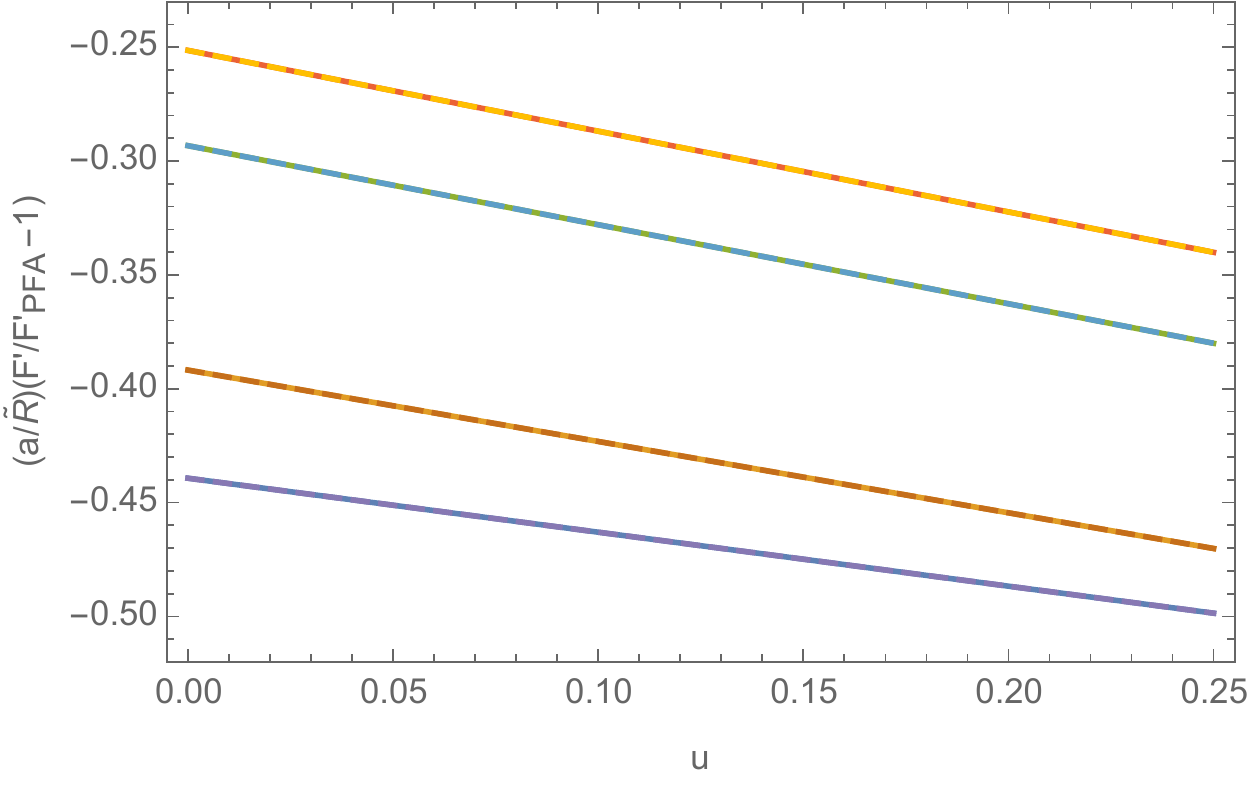}
\caption{\label{deviations2}   Beyond-PFA corrections for the gradient of the Casimir force between two gold spheres at room temperature are shown as a function of the parameter $u={\tilde R}^2/ (R_1 R_2)$ (for constant ${\tilde R}$ and $a$).  Solid lines are for   ${\tilde R}=30\;\mu$m, dashed lines for ${\tilde R}=100\;\mu$m. The four pairs of solid and dashed lines from top to bottom correspond to the four separations $a=1\mu$m, 800 nm, 400 nm and 100 nm respectively. The extreme values $u=0$ and $u=1/4$ correspond, respectively, to a sphere-plate configuration and to two spheres of equal radii.}
\end{figure}
This is clearly seen from Fig. \ref{deviations2}, where the quantity $({\tilde R}/a)(F'/F'_{\rm PFA}-1)$ is displayed versus the parameter $u$ (for constant ${\tilde R}$ and $a$). In Fig. \ref{deviations2} solid lines are for ${\tilde R}=30\;\mu$m, while dashed lines are for  ${\tilde R}=100\;\mu$m. The four pair of solid and dashed lines from top to bottom correspond to the four separations $a=1\mu$m, 800 nm, 400 nm and 100 nm respectively. We recall that the extreme values $u=0$ and $u=1/4$ correspond, respectively, to a sphere-plate and to two spheres of equal radii.  Fig. \ref{deviations2} shows  that the $u$-dependence of the deviations from PFA  is linear for the considered separations.

For small values of $a/{\tilde R}$, the quantity displayed in Fig. \ref{deviations} and in Fig. \ref{deviations2} can be basically identified with the parameter $\beta'$ that was introduced  in the sphere-plate experiment \cite{ricardo}  as a measure of the deviation of the data from PFA. In \cite{ricardo} starting from the force gradient $F'$,  an effective pressure $P^{(\rm eff)}(a,R)$ was defined  as:
\be
P^{(\rm eff)}(a,R)\equiv-\frac{F'}{2 \pi R} \;.
\ee
If the PFA Eq. (\ref{PFA2}) were exact, $P^{(\rm eff)}(a,R)=F^{(\rm pp)}(a)$.  However, the PFA is not exact, and so the authors of \cite{ricardo}  parametrized  deviations  from PFA by a  coefficient $\beta'(a)$ such that:
\be
P^{(\rm eff)}=F^{(\rm pp)}(a)\left(1+\beta' \frac{a}{R}+o(a/R)\right)\;.\label{ricardo}
\ee 
  It is clear from Eq. (\ref{ricardo}) that, up to  higher order corrections, $\beta'(a)$ coincides with  the quantity $({\tilde R}/a)(F'/F'_{\rm PFA}-1)$.
The parameter $\beta'$ was determined in \cite{ricardo} by measuring the effective pressure  $P^{(\rm eff)}$ for certain fixed sphere plate-separations using spheres of different radii, and then fitting  $P^{(\rm eff)}$ versus $1/R$ with a straight line.   
The experiment  \cite{ricardo} placed a bound $|\beta'|<0.4$ at 95 \% CL in the separation range from 150 to 300 nm.  This bound is  in substantial agreement with the theoretical prediction (see the upper curves of Fig. \ref{deviations}).  

The authors of \cite{garrett} used the same procedure to  study  deviations of their data from PFA. In particular, they assumed that  the measured forces can be parametrized as:
\be
\frac{F'}{\tilde R}=-2 \pi F^{(\rm pp)}(a)\left(1+\beta' \frac{a}{\tilde R}+o(a/{\tilde R})\right)\;,\label{garrett}
\ee 
i.e. by a function of the same form as  that of the sphere-plate system, apart from the substitution of $R$ by the effective radius ${\tilde R}$ of the two-sphere system. Importantly,  they assumed that $\beta'$ is {\it  independent of the radii of the spheres}.  Based on this assumption, the authors of \cite{garrett} tried to determine $\beta'(a)$ by doing a linear fit of $F'/{\tilde R}$ versus $1/{\tilde R}$  using for that purpose   12 measurements of $F'$. Three sets of data  were taken in sphere-plate setups (as in the experiment \cite{ricardo}), using three different spheres of radii  $R=$ 40.7 $\mu$m, 36.1 $\mu$m and 34.2 $\mu$m, while the remaining  nine data sets were taken with  nine different sphere-sphere setups,  corresponding to the different combinations of  each of three spheres with radii $R_1=$ 34.2  $\mu$m, 36.1 $\mu$m and 40.7 $\mu$m,  with each of the three spheres of  radii  $R_2=$ 29.8  $\mu$m, 38.0  $\mu$m and 46.9  $\mu$m. The measurements and the corresponding fits  were repeated for  26 values of the separation,  in the interval from 40 nm to $300$ nm.  It was found that  $\beta'=-6 \pm 27$ was within the 2$\sigma$ confidence interval of their calculated $\beta'$ for all  considered separations.

The procedure used in \cite{garrett} to determine $\beta'$ is not entirely correct, however, because the parametrization in Eq. (\ref{garrett}) misses the dependence of $\beta'$ on the parameter $u$. Indeed, by comparing Eq.(\ref{garrett}) with the small-distance expansion of the sphere-sphere force Eq. (\ref{DE2s}) one finds that $\beta'$ has the expression
\be
\beta'=- \left(\hat{\theta}(a)+u\, \hat{\kappa}(a)\right)\;.\label{betass}
\ee
This formula shows that in the two-sphere case, contrary to the assumption made in \cite{garrett}, $\beta'$ does depend on the radii of the spheres via the parameter $u$. Of course, this dependence disappears in the sphere-plate case, for which $u=0$.  The  linear dependence on $u$ of deviations from PFA is clearly visible from Fig. \ref{deviations2}. Since in \cite{garrett} the 12 combinations of radii  used  to determine $\beta'$  correspond to values of $u$ that vary from zero (for the three sphere-plate setups) to 0.2498 (corresponding to  the sphere-sphere setup with  $R_1=36.1\;\mu$m and $R_2= 38\;\mu$m), the dependence of $\beta'$ on $u$ should be considered in the data analysis.  Substituing Eq. (\ref{betass}) into Eq. (\ref{garrett}), we find that in a two-sphere system  $F'/{\tilde R}$ has the expression:
\be
\frac{F'}{\tilde R}=-2 \pi F^{(\rm pp)}(a)\left(1- \frac{a\, \hat{\theta}}{\tilde R}-\frac{a\,\hat{\kappa}}{R_1+R_2}+o(a/{\tilde R})\right)\;,\label{garrett2}
\ee
where in the second term between the brackets we used the relation $u/{\tilde R}=1/(R_1+R_2)$. This formula shows that the correct procedure to determine the coefficients  $\hat \theta$ and $\hat \kappa$,  is to make a joint 2-dimensional linear fit of $F'/{\tilde R}$ versus $1/{\tilde R}$ and $1/(R_1+R_2)$. 

The present sensitivity of the experiment \cite{garrett} is not yet sufficient   to detect the small deviations from PFA predicted by Eq. (\ref{garrett2}).  We estimate that an increase in the sensitivity by  over one  order of magnitude would be necessary for that purpose. It is hoped that future improvements of the apparatus  will achieve this goal.

\section{Conclusions}

Motivated by the recent experiment in \cite{garrett}, we have performed a precise computation of the gradient of the Casimir force between two gold spheres at room temperature.  Our computation  provides an accurate estimate of beyond PFA corrections for this system. The  semi-analytic formula  for the Casimir force  that we construct is valid for all separations and can be easily used to interpret future experiments in both the sphere-plate and sphere-sphere configurations. We have also described the correct parametrization of the corrections to PFA that should be used to carry out the data analysis in experiments using the sphere-sphere geometry. 

In our computations we modelled the god plates as  ohmic conductors (connected to charge reservoirs). In recent years it has been argued by some researchers \cite{book2} that a  better agreement with Casimir experiments is obtained if  metallic bodies are modelled as dissipationless plasmas. The main change introduced by this model is in the classical $n=0$ Matsubara term for TE polarization, which is zero within the Drude prescription, but different from zero in the plasma model. A detailed comparison between the Drude and plasma models for the sphere-plate configuration, based on a large scale numerical simulation of the scattering formula, has been reported in \cite{gert}, where it was shown that the deviations from PFA engendered by the plasma prescription  have the same qualitative behavior as the Drude model, but are slightly larger in magnitude and show a more pronounced dependence on the aspect ration $a/R$.  We plan to study the plasma prescription for the sphere-sphere case in a forthcoming work  \cite{bimontenext}.

\acknowledgments

The author thanks T. Emig, N. Graham, M. Kruger, R. L. Jaffe and M. Kardar for valuable discussions.  

\section*{APPENDIX}

\subsection{The DE expansion}

For the convenience of the reader, in this Appendix we briefly review  general properties of the DE that are useful for the present work. The DE \cite{fosco1,bimonte1,fosco2,bimonte2,fosco3}  is an asymptotic expansion that allows to compute  curvature corrections of any sufficiently {\it local} functional ${\hat {\cal F}}$ that describes the interaction between two (non-intersecting) surfaces $\Sigma_1$ and $\Sigma_2$. The idea behind the DE is intuitive. One considers that the two surfaces can be described by {\it smooth} height profiles $z=H_1(x,y)$ and $z=H_2(x,y)$, where $(x,y)$ are cartesian coordinates spanning some reference plane $\Sigma$  and $z$ is a coordinate perpendicular to  $\Sigma$ (see Fig. 1).  Since the two surfaces are non-intersecting, it can always be assumed that $H_2(x,y) < H_1(x,y)$.  At this point one considers that for surfaces of small slopes $| \nabla H_i| \ll 1,\;i=1,2$ \footnote{In fact it is sufficient that the small slope condition is satisfied only in the relevant interaction area around the point of closest approach between the two surfaces}  it should be possible to expand ${\hat {\cal F}}[H_1,H_2]$ in powers of {\it derivatives} of  increasing order of the height profiles, at least up to some order.  It is rather easy to convince oneself that for a functional ${\hat {\cal F}}[H_1,H_2]$ that is invariant under simultaneous rotations and translations of $H_1$ and $H_2$  in the reference plane $\Sigma$ (like the Casimir force between two plates made of a homogeneous and isotropic material) the most general expression of the DE valid to second order in the slopes of the surfaces is of the form:     
\begin{eqnarray}
{\hat {\cal F}}[H_1,H_2]=\int_{\Sigma}  d^2 x \left[{\hat {\cal F}}^{(\rm pp)}(H)\right. &+&\left.  \;\alpha_1(H)  (\nabla H_1)^2 \right.  \nonumber \\
+ \alpha_2(H)  (\nabla H_2)^2 \!\!&+& \!\!\alpha_{\times}(H) \nabla H_1 \cdot \nabla H_2 \nonumber\\
+  \alpha_{-}(H) \nabla H_1 \!\!& \times &\! \!\nabla H_2 \left. \right]+  \rho^{(2)}\;,\label{derexp}
\end{eqnarray}
where we set $H=H_1-H_2$ and $\rho^{(2)}$ is a remainder that becomes negligible as the local radii of curvature of the surfaces go to infinity  for fixed minimum surface-surface distance $a$. 
Note that invariance of ${\hat {\cal F}}$ under translations of $\Sigma$ in the $z$ direction implies that ${\hat {\cal F}}^{(\rm pp)}$ and the $\alpha$'s can depend only on the height difference $H$ an not on the individual heights $H_1$ and $H_2$.
It is evident that the quantity ${\hat {\cal F}}^{(\rm pp)}(a)$ in Eq. (\ref{derexp}) provides the (unit-area) interaction of two plane-parallel surfaces at distance $a$, and thus the first term on the r.h.s of Eq. (\ref{derexp}) reproduces the Derjaguin Approximation (DA)  \footnote{The DA approximation is sometimes referred to as the "exact" PFA.} for the functional ${\hat {\cal F}}$:  
\be
 {\hat {\cal F}}^{(\rm DA)} = \int_{\Sigma}  d^2 x  \;{\hat {\cal F}}^{(\rm pp)}(H)\;,
\ee
The integrals on the r.h.s. of Eq. (\ref{derexp}) that are proportional to $\alpha$'s represent curvature corrections beyond the DA, and thus we see that the DE  provides a systematic way to improve the old-fashioned DA. 
Arbitrariness in the choice of the reference plane $\Sigma$ further constraints the three coefficients $\alpha$ in Eq. (\ref{derexp}) \cite{bimonte1}. In particular,   invariance of  ${\hat {\cal F}}$ with respect to tilting of $\Sigma$ (for details, see \cite{bimonte1}) implies:
\begin{eqnarray}
&&2(\alpha_1(H)+\alpha_2(H)+\alpha_{\times}(H))+H \frac{d  {\hat {\cal F}}^{(\rm pp)} }{d H}-{\hat {\cal F}}^{(\rm pp)}=0\;,\nonumber\\
&&\alpha_{-}(H)=0\;.\label{alpharel}
\end{eqnarray}
The above relations show that, to second order in the gradient expansion, the two-surface problem actually reduces to the simpler problem of a single curved surface opposite a plane, since $\alpha_1$ and $\alpha_2$ can be determined in that case, and then $\alpha_{\times}$ follows from the first of Eqs. (\ref{alpharel}).  We now make the simplifying assumption that the  field(s) that mediate the interaction obeys the same boundary conditions on $\Sigma_1$ and $\Sigma_2$. Then 
\be\alpha_1(H)=\alpha_2(H)\equiv\alpha(H)\;.\label{onemat}
\ee
Taking advantage of Eqs. (\ref{alpharel}) and (\ref{onemat}) the DE  can then be recast in the form:
$$
{\hat {\cal F}}[H_1,H_2]= {\hat {\cal F}}^{(\rm DA)} + \int_{\Sigma}  d^2 x \; \frac{}{} \alpha(H)  (\nabla H)^2  
$$
\be
+ \frac{1}{2} \int_{\Sigma}  d^2 x \left( {\hat {\cal F}}^{(\rm pp)}-H \frac{d  {\hat {\cal F}}^{(\rm pp)}  }{d H} \right)\nabla H_1 \cdot \nabla H_2 +  \rho^{(2)}\;.\label{derexp2}
\ee
We thus see  that to second order in the slope  the interaction  ${\hat {\cal F}}$ is fully determined by knowledge of the (unit area) interaction $  {\hat {\cal F}}^{(\rm pp)}(a)$ of two parallel plates and by the single coefficient $\alpha(H)$. 

The latter coefficient can be determined by comparing the DE Eq. (\ref{derexp2}) to a {\it perturbative} expansion of  the functional ${\hat {\cal F}}[H,0]$ around flat plates $H=a + h(x,y)$ to second order in the deformation $h(x,y)$. Note that the latter perturbation requires a deformation of small amplitude  $h(x,y)/a \ll 1$,  while the DE relies on the condition that the slope of the surface be small. To second order in $h$ the perturbative expansion of ${\hat {\cal F}}$ reads:
$$
{\hat {\cal F}}[a+h({\bf x})]=A {\hat {\cal F}}^{(\rm pp)}(a)+\mu(a) {\tilde h}({\bf 0})
$$
\be
+\int \frac{d^2{\bf k}}{(2 \pi)^2}\;{\tilde G}^{(2)}(k;a)|{\tilde h}({\bf k})|^2+{\tilde \rho}^{(2)}[h]\;,
\ee
where $A$ is the surface area, ${\bf k}$ is the in-plane wave-vector,  ${\tilde h}({\bf k})$ is the Fourier transform of $h(x)$, and  ${\tilde \rho}^{(2)}[h]$  refers to higher order corrections. The function  $\alpha(H)$ can now be determined if the kernel ${\tilde G}^{(2)}(k;a)$ can be expanded to second order in $k$. Indeed, matching the expansion
\be
{\tilde G}^{(2)}(k;a)= \gamma(a)+\delta(a) k^2+o(k^2)\;,\label{green1}
\ee
to Eq. (\ref{derexp2}) one finds:
\be
 {\hat {\cal F}}^{(\rm pp)}\!\frac{}{}'(a)=\mu(a)\;,\;\;{\hat {\cal F}}^{(\rm pp)}\!\frac{}{}''(a)=2 \gamma(a)\;,\;\;\;\alpha(a)=\delta(a)\;,\label{green2}
\ee
where a prime denotes a derivative with respect to $a$. The above Equation shows that a {\it necessary} condition for existence of the second order DE is existence of the Taylor expansion of the perturbative kernel ${\tilde G}^{(2)}(k;a)$ to second order in the in-plane momentum. Indeed, it can be shown that the DE can be formally recovered by an (infinite) resummation of the perturbative series for small in-plane momenta \cite{fosco3}.

Whenever applicable, the DE has been successfully used to compute curvature corrections beyond the PA in various problems involving interactions among gently curved surfaces. In the context of Casimir physics, it was used in \cite{fosco1} to  compute curvature corrections to the zero temperature Casimir energy for a scalar field obeying Dirichlet (D) boundary conditions (bc) in the  sphere-plate and cylinder-plate  geometries. The zero temperature Casimir problem for the em field with perfect conductor (PC) bc, as well as a scalar field obeying Neumann bc, or  mixed DN bc (i.e. D bc on one surface and N on the other), was studied in \cite{bimonte1} for two spheres and for two inclined  cylinders. The curvature corrections obtained in the latter work  for the em field with PC bc in the sphere-plate and sphere-sphere geometries  were subsequently  confirmed in \cite{bordagteo,teo2} by working out a rigorous small-distance expansion of the scattering formula. The experimentally important case of the Casimir interaction between gold sphere and plate at finite temperature was instead studied in \cite{bimonte2}. Even in this case, the results obtained by the DE were later shown to be in agreement with the small-distance expansion of the scattering formula \cite{teogold}. Curvature corrections obtained by the DE have also been found to be in agreement 
with the small distance expansion of the rare exact Casimir energies in non planar geometries that have been discovered so far,  i.e. in the cases of two Drude or D spheres in the classical limit \cite{bimonteex1}, and for two three-spheres with D or PC bc in four euclidean dimensions \cite{euclidean}.    
The DE has been also used to study curvature effects in the Casimir-Polder interaction of a particle with a gently curved surface \cite{CPbimonte1,CPbimonte2}, 
and to estimate the shifts of the rotational levels of a diatomic molecule due to its van der Waals interaction with a curved dielectric surface \cite{CPbimonte3}. In a non Casimir context, the DE hase been also used to compute curvature corrections to the scattering amplitude for an em wave impinging on a curved surface \cite{bimonteref} and to the electrostatic interaction among two curved plates \cite{foscoel}.
   

\subsection{Computing the leading curvature correction to the force gradient. }


The  small-slope approximation of the interaction energy ${\hat {\cal F}}$ provided by Eq. (\ref{derexp2}) still involves a surface integral over $\Sigma$ of functions depending on the height profiles of the surfaces. As such, 
Eq.  (\ref{derexp2}) is not very convenient for a practical use. A better route is to expand Eq. (\ref{derexp2}) in powers of the small parameter $a/R$, where $R$ is the characteristic radius of curvature of the surfaces. The leading order of this expansion will reproduce the standard PFA, while in the next order it shall provide us with the desired curvature correction beyond the PFA.  
We shall carry out this expansion not directly for the energy ${\hat {\cal F}}$, but rather for the gradient of the force ${\hat F}'=-{\hat {\cal F}}''$, which is the quantity that was measured in the experiment \cite{garrett}. Moreover, we shall restrict attention to the sphere-sphere system, which is again the geometry used in \cite{garrett}.

According to Eq. (\ref{derexp2}), the formula for the force gradient ${\hat F}'$ can be split as
\be
{\hat F}'={\hat { F'}}^{(\rm DA)}+I_2+I_3\;,
\ee
where
\be
{\hat { F'}}^{(\rm DA)}=\int_{\Sigma}  d^2 x  \;{\hat { F'}}^{(\rm pp)}(H)
\ee
and we set
\begin{eqnarray}
&I_2& =\frac{1}{2} \int_{\Sigma}  d^2 x\; (H {\hat {\cal F}}^{(\rm pp)}\!\frac{}{}'')'\;\nabla H_1 \cdot \nabla H_2\;,
\nonumber \\
&I_3&=- \int_{\Sigma}  d^2 x \;  \alpha''(H) \;  (\nabla H)^2\;.
\end{eqnarray}
We consider proximity forces that decay rapidly wih the distance, like the Casimir force. For forces of this nature, the  interaction among the surfaces is localized within a small area, typically of radius $\rho \sim \sqrt{a {\tilde R}}$,  around the point of closest approach.  Under such circumstances,  it is legitimate to take the Taylor expansion of the  height profiles $H_1(x,y)$ and $H_2(x,y)$ of the two spheres around their tips, that we imagine placed at $x=y=0$. Since the position of the reference plane $\Sigma$ in Fig. 1 is immaterial, we are free to take for $\Sigma$ the tangent plane to the sphere of radius $R_2$, passing through the sphere tip. Then: 
\begin{eqnarray}
H_1(x,y) &=& a+\frac{r^2}{2R_1}+\frac{r^4}{8 R_1^3}+\dots\;,\nonumber \\
H_2(x,y) &=&-\frac{r^2}{2R_2}-\frac{r^4}{8 R_2^3}+\dots\;,
\end{eqnarray}
where $r^2=x^2+y^2$.  To evaluate the integrals $I_j$ it is convenient to introduce polar coordinates $(r, \theta)$ in the $(x,y)$ plane, and then substitute $r$ by the dimensionless quantity $\xi=r^2/a {\tilde R}$.  An essential property of the integrals $I_j$ is that they involve derivatives of certain functions (i.e. ${\hat {\cal F}}^{(\rm pp)}$ and $\alpha$) of the height difference $H$ with respect to the separation $a$. These derivatives  can be converted into derivatives with respect to $\xi$, using the identity
\begin{eqnarray}
U'&=&U_{ ,\xi} \left(H_{,\xi}\right)^{-1} \nonumber \\
&=&2\frac{U_{ ,\xi} }{a}  \left[1-\frac{\xi}{2} \left(\frac {a{\tilde R}^2}{R_1^3}+\frac{a{\tilde R}^2}{R_2^3} \right) +o(a/{\tilde R})\right]\;,\label{dera}
\end{eqnarray}
(comas denote derivatives) which holds for any function $U$ of $H$. We are ready now to take the  small-distance expansion of ${\hat F}'$. We start from ${\hat { F'}}^{(\rm DA)}$. 
Using Eq. (\ref{dera}), and omitting corrections of order $o(a/{\tilde R})$ we find 
$$
{\hat { F'}}^{(\rm DA)}\!\!\!=2\pi {\tilde R} \!\!\int_0^{\infty} \!\!\!\!d \xi \,{\hat {F}}^{(\rm DA)}_{,\xi}\!\left[1-\frac{\xi}{2} \left(\!\frac{a{\tilde R}^2}{R_1^3}+\frac{a{\tilde R}^2}{R_2^3}\! \right)\right] 
$$
$$
=-2 \pi {\tilde R} {\hat F}^{(\rm pp)}(a)-a\, \pi  \left(\frac{{\tilde R}^3}{R_1^3}+\frac{{\tilde R}^3}{R_2^3} \right)
 \int_0^{\infty} d \xi \;{\hat {\cal F}}^{(\rm pp)}\! \frac{}{}'
$$
\be
=-2 \pi {\tilde R} {\hat F}^{(\rm pp)}(a)+ 2\pi  \left(\frac{{\tilde R}^3}{R_1^3}+\frac{{\tilde R}^3}{R_2^3} \right) {\hat {\cal F}}^{(\rm pp)}(a)\;.\label{first}
\ee
The first term on the last line of Eq. (\ref{first}) coincides with the standard PFA for the force gradient (see Eq. (\ref{PFA2})), while its second term represents a curvature correction. 

By following an analogous procedure for $I_2$, and again omitting higher order terms, we obtain:
\begin{eqnarray}
&I_2 & = - \pi\frac{ a {\tilde R}^2}{R_1 R_2} \int_0^{\infty} d \xi\,(H {\hat {\cal F}}^{(\rm pp)}\!\frac{}{}'')_{,\xi}\,\xi
\nonumber \\
&=&
 \pi \frac{a {\tilde R}^2}{R_1 R_2} \int_0^{\infty} d \xi\,(H {\hat {\cal F}}^{(\rm pp)}\!\frac{}{}'')=\frac{2\pi {{\tilde R}^2}}{R_1 R_2} \int_0^{\infty} d \xi\,H ({\hat {\cal F}}^{(\rm pp)}\!\frac{}{}')_{,\xi}
\nonumber \\
&=& \frac{2\pi {{\tilde R}^2}}{R_1 R_2} a {\hat F}^{(\rm pp)}(a)-\frac{2\pi {{\tilde R}^2}}{R_1 R_2} \int_0^{\infty} d \xi\, {\hat {\cal F}}^{(\rm pp)}_{,\xi} \nonumber \\
&=& \frac{2\pi {{\tilde R}^2}}{R_1 R_2} \left[a {\hat F}^{(\rm pp)}(a) +{\hat {\cal F}}^{(\rm pp)}(a)\right]\;.\label{I2}
\end{eqnarray}
Finally, for $I_3$ we obtain:
\be
I_3 = - 4 \pi \int_0^{\infty} d \xi\, \alpha_{,\xi \xi} \,\xi =-4 \pi \alpha(a)\;.\label{I3}
\ee
Upon combining Eqs. (\ref{first}-\ref{I3}), after simple algebraic transformations, we obtain the following small-distance expansion of the force gradient, correct up to terms of order $o(a/{\tilde R})$:
\begin{eqnarray}
{\hat F}' &=&-2 \pi {\tilde R} {\hat F}^{(\rm pp)}(a)+ 2 \pi \left[{\hat {\cal F}}^{(\rm pp)}(a) -2 \alpha(a)\right]
\nonumber \\
&+& 2 \pi u \left[ a {\hat F}^{(\rm pp)}(a)-2 {\hat {\cal F}}^{(\rm pp)}(a) \right]\;
\nonumber \\
&\equiv& -2 \pi {\tilde R} {\hat F}^{(\rm pp)}(a) \left[1- \left(\tilde{\theta}(a)+u \kappa(a) \right) \frac{a}{\tilde R} \right]\;, \label{DEformula}
\end{eqnarray}
where the coefficients $\tilde \theta(a)$ and $\kappa(a)$ are
\begin{eqnarray}
{\tilde \theta} &=&  \frac{{\hat {\cal F}}^{(\rm pp)}(a) -2 \alpha(a)}{a {\hat F}^{(\rm pp)}(a)}\;,\label{thetacoe}
\\
\kappa(a) &=&  1-2 \frac{{\hat {\cal F}}^{(\rm pp)}(a)}{ a {\hat F}^{(\rm pp)}(a)} \;.\label{kappacoe}
\end{eqnarray}

\end{document}